\documentclass[prc,aps,%12pt,
showpacs,nofootinbib,%
tightenlines,%
superscriptaddress,twocolumn,%
floatfix]{revtex4}
\usepackage{latexsym}
\usepackage{amsmath,amssymb, mathrsfs}
\usepackage[latin1]{inputenc}
\usepackage[dvips]{graphicx}
\usepackage[dvips]{hyperref}
\newcommand{\sd}{{\sigma_\textrm{D}}}
\newcommand{\stot}{{\sigma_\textrm{tot}}}

\newcommand{\xbj}{{x}}

\newcommand{\qs}{{Q_\mathrm{s}}}
\newcommand{\qsprime}{{Q'_\mathrm{s}}}

\newcommand{\as}{{\alpha_{\mathrm{s}}}}

\newcommand{\rt}{{\mathbf{r}_T}}
\newcommand{\xt}{{\mathbf{x}_T}}
\newcommand{\bt}{{\mathbf{b}_T}}
\newcommand{\yt}{{\mathbf{y}_T}}
\newcommand{\zt}{{\mathbf{z}_T}}

\newcommand{\nc}{{N_\mathrm{c}}}
\newcommand{\cf}{{C_\mathrm{F}}}

\newcommand{\gev}{\textrm{ GeV}}

\newcommand{\ra}{R_A}
\newcommand{\rp}{R_p}

\newcommand{\Aavg}[1]{\left\langle #1 \right\rangle_\textrm{N}}
\newcommand{\Aavgg}[1]{\left\langle #1 \right\rangle_{\textrm{N}}^2}
\newcommand{\ampli}{{\mathcal{N}}}
\newcommand{\amplibindep}{{N}}

\newcommand{\nr}[1]{(\ref{#1})} 
\newcommand{\ud}{\, \mathrm{d}}
\newcommand{\fig}{Fig.~}
\newcommand{\figs}{Figs.~}
\newcommand{\eq}{Eq.~}

\newcommand{\eqs}{Eqs.~}

\newcommand{\sigmaa}{{ \sigma^A_\textrm{dip} }}
\newcommand{\sigmap}{{ \sigma^\textrm{p}_\textrm{dip} }}

\newcommand{\dsigmap}{{\frac{\ud \sigma^\textrm{p}_\textrm{dip}}{\ud^2 \bt}}}
\newcommand{\dsigmaa}{{\frac{\ud \sigma^A_\textrm{dip}}{\ud^2 \bt}}}

\newcommand{\dsigma}{{\frac{\ud \sigma_\textrm{dip}}{\ud^2 \bt}}}

\newcommand{\dsigmaadj}{{\frac{\ud \tilde{\sigma}_\textrm{dip}}{\ud^2 \bt}}}

\newcommand{\fd}{{F_2^{\textrm{D}}}}
\newcommand{\fdp}{{{F_{2p}^\textrm{D}}}}
\newcommand{\fda}{{{F_{2A}^\textrm{D}}}}
\newcommand{\fdl}{{{F_{L,q\bar{q}}^\textrm{D}}}}
\newcommand{\fdt}{{{F_{T,q\bar{q}}^\textrm{D}}}}
\newcommand{\fdqqbg}{{{F_{T,q\bar{q}g}^\textrm{D (GBW)}}}}
\newcommand{\fdms}{{{F_{T,q\bar{q}g}^\textrm{D (MS)}}}}
\newcommand{\xpom}{{x_\mathbb{P}}}

\begin{document}

\title{
Nuclear enhancement and suppression of diffractive structure functions at high energies
}

\preprint{arXiv:0805.4071 [hep-ph]}
\preprint{IPhT-T08/089}

\author{H. Kowalski}
\affiliation{Deutsches Elektronen-Synchrotron DESY, 22607 Hamburg, Germany}
\author{T. Lappi}
\affiliation{Institut de Physique Th\'eorique,
B\^at. 774, CEA/DSM/Saclay, 91191 Gif-sur-Yvette, France}
\author{C. Marquet}
\affiliation{Institut de Physique Th\'eorique,
B\^at. 774, CEA/DSM/Saclay, 91191 Gif-sur-Yvette, France}
\affiliation{Department of Physics, Columbia University, New York, NY 10027, USA}
\author{R. Venugopalan}
\affiliation{Physics Department, Brookhaven
National Laboratory, Upton, NY 11973, USA}

\begin{abstract}
We compute diffractive structure functions for both protons and 
nuclei in the framework of Color Glass Condensate models with impact parameter
 dependence. These models have previously been shown to provide good agreement
 with inclusive $F_2$ measurements and exclusive vector meson measurements at
 HERA.  For nuclei, they provide good (parameter free) agreement with the 
inclusive $F_2$ data. We demonstrate good agreement of our computations 
with HERA measurements on inclusive diffraction. We extend our analysis 
to nuclei and predict the pattern of enhancement and suppression of the 
diffractive structures functions that can be measured at an Electron Ion 
Collider. We discuss how the impact parameter dependence crucially affects
 our analysis, in particular for large invariant masses at fixed $Q^2$.
\end{abstract}


\maketitle

\section{Introduction}

The discovery that about 15\%
of Deeply Inelastic Scattering (DIS) events at HERA are diffractive 
events~\cite{Derrick:1993xh,Ahmed:1994nw} has focused attention on the nature of 
hard diffractive
 scattering in QCD at collider energies. In particular, it is observed that the ratio
 of diffractive to inclusive cross-sections is nearly independent of the 
energy~\cite{Adloff:1997sc,Breitweg:1997aa}, and that both cross-sections display geometric 
scaling~\cite{Stasto:2000er,Marquet:2006jb,Gelis:2006bs}. It was noted some
 time ago that these features could be understood~\cite{Golec-Biernat:1999qd}
 in a simple model, the Golec-Biernat--Wusthoff (GBW) dipole 
model~\cite{Golec-Biernat:1998js}, which incorporated the physics of QCD 
saturation at high energies. The saturation of parton densities is due to non-linear
 multi--parton effects such as recombination and screening which deplete the gluon 
density at small $x$~\cite{Gribov:1984tu}. These non-linear effects are large for 
modes in the hadronic wavefunctions with transverse momenta $k_\perp \lesssim \qs$, 
where $\qs(x)$, appropriately called the saturation scale, is a scale generated by the 
multi-parton dynamics. Though the Golec-Biernat--Wusthoff model explains qualitative 
features of the inclusive and diffractive data, it fails in detailed comparisons to the
 data. This is primarily because, except for the quark masses,
 the model does not contain geometric scaling violations, 
such as for instance the bremsstrahlung limit of perturbative QCD (pQCD) that applies to
 small dipoles of transverse size $r \ll 1/\qs(x)$. 

The GBW model is significantly improved by including the appropriate DGLAP behavior 
for dipoles with small transverse sizes~\cite{Bartels:2002cj}. This  ``DGLAP improved''
 dipole model arises naturally in the classical effective theory of the Color Glass
 Condensate (CGC)~\cite{McLerran:1994ni,McLerran:1994ka,McLerran:1994vd}. 
For instance, in the CGC~\cite{Iancu:2003xm,Weigert:2005us}, one obtains the well known 
expression~\cite{Mueller:1989st,Nikolaev:1990ja} for the inclusive virtual photon 
hadron cross section
\begin{equation}\label{eq:sigmatot}
\sigma^{\gamma^*p}_{L,T}
= \int\! \ud^2 \rt \int_0^1 \! \ud z \left| \Psi^{\gamma^*}_{L,T}
\right|^2 
\int \! \ud^2 \bt \dsigmap  \, ,
\end{equation}
where $\left| \Psi_{L,T}^{\gamma^*}(\rt,z,Q) \right|^2$
represents the probability for a virtual photon to produce a quark--anti-quark pair 
of size $r = |\rt|$ and $\dsigmap(\rt,\xbj,\bt)$ denotes the 
\emph{dipole cross section} for this pair to scatter off the target at an 
impact parameter $\bt$. The former is well known from QED, while the latter 
represents the dynamics of QCD scattering at small $x$. In the ``classical''
limit without high energy evolution effects, the dipole cross section 
can be written as
\begin{equation}
\dsigmap
 = 2\,\left[ 1 - \exp\left(- r^2  F(\xbj,r) T_p(\bt)\right) 
\right],
\label{eq:BEKW}
\end{equation}
where $T_p(\bt)\sim\exp(-\frac{b^2}{2 B_{\rm G}})$ is the impact parameter profile
 function in the proton, normalized as 
$\int d^2 \bt \,T_p(\bt) = 1$ and $F$ is proportional to the 
DGLAP evolved gluon distribution~\cite{Bartels:2002cj}
\begin{equation}
F(\xbj,r^2) = \frac{ \pi^2 }{2 \nc} \as \left(\mu_0^2 + \frac{C}{r^2} \right) 
\xbj g\left(\xbj,\mu_0^2 + \frac{C}{r^2} \right)  .
\label{eq:BEKW_F}
\end{equation}
In general, the dipole cross section can range from $0$ in the
$r \to 0$ color transparency limit to $2$, the maximal unitarity bound.
The saturation scale $\qs$ characterizes the qualitative change
between these regimes; following~\cite{Kowalski:2007rw} 
we shall define here $\qs$ as the solution  of 
\begin{equation}\label{eq:defQs}
\dsigma(\xbj , r^2 = 1/\qs^2(\xbj,\bt)) = 2(1-e^{-1/4})
\end{equation}
Note that our definition is completely model independent and can be applied to any 
sensible parametrization of the dipole cross section.
The definition \nr{eq:defQs} applied to a Gaussian dipole cross section gives the 
saturation scale of the GBW model,
but it differs slightly from the convention in 
Ref.~\cite{Kowalski:2003hm}, where the saturation criterion was taken as
$\dsigma = 2(1-e^{-1/2})$ at $r^2 = 2/\qs^2$. 
The dipole cross section in \eq\nr{eq:BEKW} was 
implemented in the impact parameter saturation model (IPsat)~\cite{Kowalski:2003hm} 
where the parameters $\mu_0,$ $C,$ and $B_{\rm G}$ (as well as two other parameters 
characterizing the  initial condition for the DGLAP evolution) are fit to
reproduce the HERA data on the  inclusive  structure function $F_2$.

The form of the IPsat dipole cross section in \eq\nr{eq:BEKW} is applicable
when leading logarithms in $Q^2$ dominate over leading logarithms in $x$. 
At very small $x$, quantum evolution in the CGC describes both the dilute 
bremsstrahlung limit of linear small $x$ evolution as well as nonlinear RG
evolution at high parton densities~\cite{Iancu:2003xm}.
The essential dynamics of this small $x$ evolution
are combined with a more realistic 
$b$-dependence
in the bCGC model~\cite{Iancu:2003ge,Kowalski:2006hc}. The model is formulated in terms
of an explicit $x$-dependent saturation scale that we shall denote by 
$\qsprime$ to distinguish it from our model independent definition \nr{eq:defQs}.
The dipole cross section has the form 
\begin{align}
\dsigmap &= 2\,  {\cal N}_0 \left(\frac{r\qsprime}{2}\right)^{2\left(\gamma_s + 
{1\over \kappa \lambda Y}\ln\left(\frac{2}{r\qsprime}\right)\right)} &  \textrm{ for } & 
r\qsprime \leq 2\nonumber \\ 
&= 2 - 2\exp\left(-A\ln^2\left(Br\qsprime\right)\right) & \textrm{ for } & r\qsprime > 2 \, .
\label{eq:b-CGC}
\end{align}
The coefficients $A$ and $B$ in the second line of this equation can be determined uniquely 
from the condition that $\dsigmap$ and its first 
derivative with respect to $r\qsprime$ are continuous across $r\qsprime =2$. 
Here $Y = \ln(1/\xbj)$ is the rapidity, while $\gamma_s =0.628$ and $\kappa = 9.9$ (which quantifies the geometric scaling violations in \eqref{eq:b-CGC}) are obtained from the leading logarithmic BFKL dynamics \cite{Iancu:2002tr}.
The impact parameter dependence of the proton saturation scale is introduced into the 
bCGC model in the form
\begin{equation}
\qsprime(\xbj,b) = \left(\frac{x_0}{\xbj}\right)^{\frac{\lambda}{2}}\left[\exp\left( 
- b^2/2 B_{\rm CGC}\right)\right]^{\frac{1}{2\gamma_s}} \gev \, .
\label{eq:qs-bCGC}
\end{equation}
After choosing ${\cal N}_0=0.7,$ the parameters $\lambda$, $x_0$ and $B_{\rm CGC}$ are
 fit to the data. We will discuss the impact parameter dependence of the dipole 
cross-section and the saturation scale further in the next section.
We must emphasize that 
the saturation scale $\qs$ is conceptually the same as $\qsprime$ and 
their numerical values are of the same order, but we differentiate between them in order
to maintain our model independent definition \nr{eq:defQs} and the original parametrization
of the bCGC model.

Both the IPsat model and the bCGC model provide excellent fits 
to a wide range of HERA data for $x \leq 0.01$~\cite{Kowalski:2006hc,Forshaw:2006np}. In Ref.~\cite{Kowalski:2007rw}, we 
discussed the possibility that DIS off nuclei can distinguish respectively between these  
``classical CGC'' and ``quantum CGC'' motivated models. 
Our discussion in that paper addressed the $A$ dependence of the nuclear saturation scale 
and fits to the available nuclear DIS inclusive data. We also addressed elastic scattering
 of $q\bar{q}$ dipoles of nuclei but our analysis was incomplete because a more complete 
picture of diffraction requires that one consider the diffractive scattering off nuclei 
of higher Fock states as well --- at least of the $q\bar{q} g$ Fock state. This 
shortcoming is addressed in the present paper. We note that there have been several 
discussions of diffractive scattering on nuclei in the 
literature~\cite{Nikolaev:1991et,Nikolaev:1995xu,Frankfurt:1991nx,Frankfurt:2001av,Frankfurt:2002kd,Gotsman:1999vt,Levin:2001et,Levin:2002fj,Kopeliovich:1999am,Kopeliovich:2006bm,Kugeratski:2005ck}.
We will later compare and contrast our results with those studies that overlap with ours. 

The paper is organized as follows. In the next section, we will introduce the 
 kinematics for diffractive deep inelastic scattering and the relevant
 formulae to compute diffractive structure functions in the dipole model.
  We will discuss how the $q\bar{q}$ and $q\bar{q}g$ dipoles, 
the dominant contributions at present collider energies, contribute to the
 diffractive cross-section. Particular attention will be paid to the impact 
parameter dependence of the cross-sections. In Sec.~III, we will discuss a comparison 
of the different CGC based models to the HERA 
diffractive data. This analysis is extended to nuclei in Sec.~IV. The ``breakup'' and
 ``non breakup'' events discussed in Ref.~\cite{Kowalski:2007rw} for 
$q\bar{q}$ dipoles are applied to the $q\bar{q}g$ dipoles as well. We study the nuclear 
enhancement and suppression of diffractive cross-sections in the 
different models and assess their predictive power. In the final section we compare our 
results with those existing in the literature and outline future research on this topic. 

\section{Computing the diffractive structure function}\label{sec:f2d3}

\begin{figure}
\includegraphics[width=0.45\textwidth]{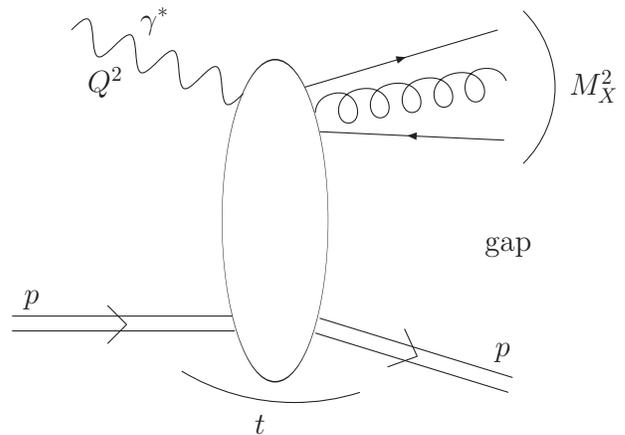}
\caption{ Kinematics of diffractive DIS.
}
\label{fig:DIS-kin}
\end{figure}

In the dipole picture of diffractive DIS, the virtual photon fluctuates into a colorless
 parton Fock state, which interacts elastically with the 
proton or nucleus. In the final state, the latter remains intact, while (for inclusive 
diffraction), the parton Fock state fragments into hadrons with an 
invariant mass $M_X$. In addition to the usual DIS invariants $x$ and $Q^2$, inclusive 
diffractive scattering can be fully characterized by two other invariants $\beta$ and 
$t$. Here $t= -(P - P^\prime)^2$, where $P^\mu$ and ${P^\prime}^\mu$ denote the 
four-momenta of the incoming and outgoing hadron, and $\beta = Q^2/(Q^2 + M_X^2)$,
 with $0 <\beta < 1$. 
Note that if $Y = \ln(1/\xbj)$ is the relative rapidity 
between the electron and the hadron, $\ln(1/\beta)$ is the rapidity interval between 
the electron and the hadronic fragments $X$ of the Fock state. The rapidity gap 
between $X$ and the proton or nucleus is  expressed as $Y_{\mathbb{P}} = \ln(1/\xpom)$;
 therefore, by definition, $Y = \ln(1/\beta) + Y_{\mathbb{P}}$, or $\xpom = \xbj/\beta$. 
The kinematics is illustrated in \fig\ref{fig:DIS-kin}.

The parton Fock state could be a $q\bar{q}$ dipole or higher Fock states which 
involve emission of one or more gluons. Naively, these higher 
Fock states are suppressed by higher powers of $\as$; however, in the limits 
of large $Q^2$ or small $\beta$, gluon emissions are accompanied by large 
logarithms in $Q^2$ and $\beta$ respectively that compensate the powers of 
the coupling. These factors then have to be resummed to obtain the diffractive
 cross-sections in the appropriate kinematic domains. The powers of
$\as \ln 1/\beta$ can in principle be resummed in the dipole picture by  
using the Kovchegov-Levin 
equation \cite{Kovchegov:1999ji} in the mean-field approximation, and the approach 
of \cite{Hatta:2006hs} beyond. In practice one must require $\xpom < 0.01$ for
the rapidity gap to be large enough to be clearly detectable and to be in
the domain of validity of the dipole model. Thus in the kinematical regime of HERA 
(and realistic  future DIS experiments) $\ln 1/ \beta$ is not very large 
and we shall only take into account the  contribution of the $q\bar{q}$ 
and $q\bar{q}g$ components in the present work. 

In diffractive scattering, contrary to inclusive scattering, one
is really computing the square of an amplitude --- 
with interference terms. As a consequence, the 
square of the photon wave function cannot be factorized from the cross section.
Introducing the auxiliary notation
\begin{equation}
\Phi_n = \int \! \ud^2 \bt 
\left[ \int_0^\infty \!\!\!\!\!\! \ud r 
r K_n(\varepsilon r) J_n(kr) \dsigma(\bt,r,\xpom) \right]^2
\label{eq:phi-n}
\end{equation}
the $q\bar{q}$ components of the diffractive structure function. 
can be expressed as
\begin{multline}
\xpom \fdt(\xpom,\beta,Q^2) =
% \\
 \frac{\nc Q^4}{16 \pi^3 \beta}\sum_{f}e_f^2 
\int_{z_0}^{1/2}\ud z z(1-z) 
\\ 
\times \left[
\varepsilon^2(z^2 + (1-z)^2) \Phi_1 + m_f^2 \Phi_0
\right]
\end{multline}
\begin{multline}
\xpom \fdl(\xpom,\beta,Q^2) = 
% \\
\frac{\nc Q^6}{4 \pi^3 \beta}\sum_{f}e_f^2 
\int_{z_0}^{1/2}
\ud z z^3 (1-z)^3 
\Phi_0
\end{multline}
with $\varepsilon^2 = z(1-z)Q^2 + m_f^2$, $k^2=z(1-z)M_X^2-m_f^2$
and $z_0 = \left(1-\sqrt{1-4m_f^2/M_X^2} \right)/2$.
In principle the measurable $|t|$ is bounded by a very small kinematical lower
bound $ |t| > (\xpom m_N)^2$  and both upper and lower experimental limits.
Here we have, however, performed the $t$ integration from $-\infty$ to $0,$ 
because this enables us write the above expressions in impact parameter space.
Since the
decrease of the structure functions with $t$ is (close to an) exponential,
this approximation is reasonable. Note that we allow for an arbitrary impact 
parameter dependence, and thus there is no ``diffractive slope'' as a separate parameter.

For the $q\bar{q}g$ component of the structure function, there are essentially two 
approaches that have been used. The one derived in the
large $Q^2$-limit in Ref.~\cite{Wusthoff:1997fz} gives 
\begin{multline}\label{eq:gbw}
\xpom \fdqqbg(\xpom,\beta,Q^2) = 
\\
\frac{\as\beta}{8\pi^4}\sum_{f}e_f^2 
\int \ud^2 \bt
\int_0^{Q^2} \ud k^2 
\int_{\beta}^{1} \ud z \Bigg\{
\\
k^4 \ln \frac{Q^2}{k^2}
\left[\left(1\!-\!\frac{\beta}z\right)^2+\left(\frac{\beta}z\right)^2\right]
\\
\times \bigg[\int_0^\infty \ud r r 
 \dsigmaadj(\bt,\rt,\xpom) 
\\
K_2(\sqrt{z}k r) J_2(\sqrt{1-z}kr)
\bigg]^2
\Bigg\}.
\end{multline}
In this limit, the $q\bar{q}g$-system is an adjoint representation $gg$ dipole, 
therefore, as in Ref.~\cite{Marquet:2007nf}, we use here the dipole cross section in the
adjoint representation, denoted by 
\begin{equation}
\dsigmaadj = 2 \left[1-\left(1-\frac{1}{2} \dsigma \right)^2 \right].
\label{eq:dip-adjoint}
\end{equation}
This large $\nc$ expression, where the $gg$-system is treated as two 
fundamental representation $q\bar q$ dipoles, is consistent with the 
Balitsky-Kovchegov~\cite{Balitsky:1995ub,%
Kovchegov:1999yj,Kovchegov:1999ua}
mean field treatment of small $x$ evolution. The same is true of the $\beta \to 0$-limit
(\eq\nr{eq:ms}) below, where the $q\bar qg$-system is also treated as two 
fundamental representation $q\bar q$ dipoles, but this time with different sizes.
We will refer to \eq\nr{eq:gbw} as the ``GBW'' $q\bar{q}g$-component 
(referring to \cite{Golec-Biernat:1999qd}).
 Note that the form \nr{eq:dip-adjoint} differs from the one used in 
Ref.~\cite{Golec-Biernat:1999qd}, where the adjoint dipole does not have the 
right saturation limit. 

The other well studied case is the $\beta \to 0$ limit~\cite{Bartels:1999tn,%
Kopeliovich:1999am,Kovchegov:2001ni,Munier:2003zb,Marquet:2004xa,GolecBiernat:2005fe}.
In this limit (the notation ``MS'' below refers to the authors of~\cite{Munier:2003zb}) 
the structure
function again factorizes into the product of a photon wave function and the cross 
section for  the $q\bar{q}g$-system to interact elastically (diffractively) 
with the target: 
\begin{multline}\label{eq:ms}
\xpom \fdms(\xpom,\beta=0,Q^2)
=
\frac{\cf \as Q^2 }{4 \pi^4 \alpha_{\mathrm{em}}}
\int \ud^2 \rt \int_0^1 \! \ud z 
\\
\left| \Psi^{\gamma^*}_T(r,Q,z)\right|^2
\int \ud^2 \bt
 A(r,\xpom,\bt),
\end{multline}
with
\begin{multline}\label{eq:MS-amplitude}
A(r,\xpom,\bt) = 
\int \ud^2 \rt' \frac{\rt^2}{\rt'^2(\rt-\rt')^2}
\bigg[
\ampli(\rt') 
\\
+ \ampli(\rt-\rt') 
- \ampli(\rt)
-\ampli(\rt') \ampli(\rt-\rt')
\bigg]^2
\end{multline}
and $\ampli(r,\xpom,\bt)=\dsigma(r,\xpom,\bt)/2.$ The coupling constant $\as$ in
 \eqs\nr{eq:gbw} and \nr{eq:ms} is treated here as a constant
free parameter (independent of the DGLAP evolution momentum
scale in the IPsat model). A more thorough study of the running coupling effects in this
 problem is an interesting question that is out of the scope of the present work.

\begin{figure}
\includegraphics[width=0.45\textwidth]{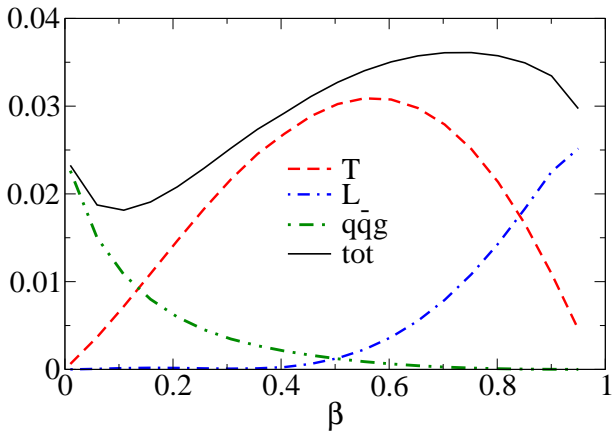}
\caption{
$\beta$-dependence of the different contributions to the proton
diffractive structure function at $Q^2=5\gev^2$ and $\xpom = 10^{-3}$.
}
\label{fig:f2d3pbetadep}
\end{figure}

Depending on the mass of the diffractive system $M_X$ or, equivalently, $\beta$,
the diffractive structure function is dominated by either the $q\bar{q}$ or $q\bar{q}g$ 
Fock 
states. (See Ref.~\cite{Bartels:1998ea} for a general argument of the $\beta$ dependence.)
Specifically, in the limit $\beta \to 1$, the dominant component, $\fdl$ is a longitudinally
polarized $q\bar{q}$ system. At intermediate $\beta\sim 0.5$ the dominant component is 
a transversally polarized $q\bar{q}$ denoted here by $\fdt$. In the limit $\beta\to 0$
the invariant mass of the diffractive system is large, and this large phase space is filled
by radiation of additional gluons, each of them being suppressed by $\as$. 
This structure is illustrated in \fig\ref{fig:f2d3pbetadep}.

In Ref.~\cite{Marquet:2007nf}, it was shown that the $\beta=0$ limit of \eq\nr{eq:gbw}, 
at $Q^2\rightarrow \infty$, approaches the result from \eq\nr{eq:ms}. This therefore 
suggests 
the following interpolation formula between the two limits~\cite{Marquet:2007nf}:
\begin{multline}
\xpom F_{T,q\bar{q}g}(\xpom,\beta,Q^2) =
\\ 
\frac{\xpom \fdqqbg(\xpom,\beta,Q^2) \times \xpom \fdms(\xpom,Q^2) }
{\xpom \fdqqbg(\xpom,\beta=0,Q^2)} .
\label{eq:Cyrille-interp}
\end{multline}
We also note the work in Refs.~\cite{Gotsman:1999vt,Levin:2001pr,Levin:2002fj},
which uses yet another prescription whose relation to ours is not very transparent.

\section{Impact parameter dependence}

\begin{figure}
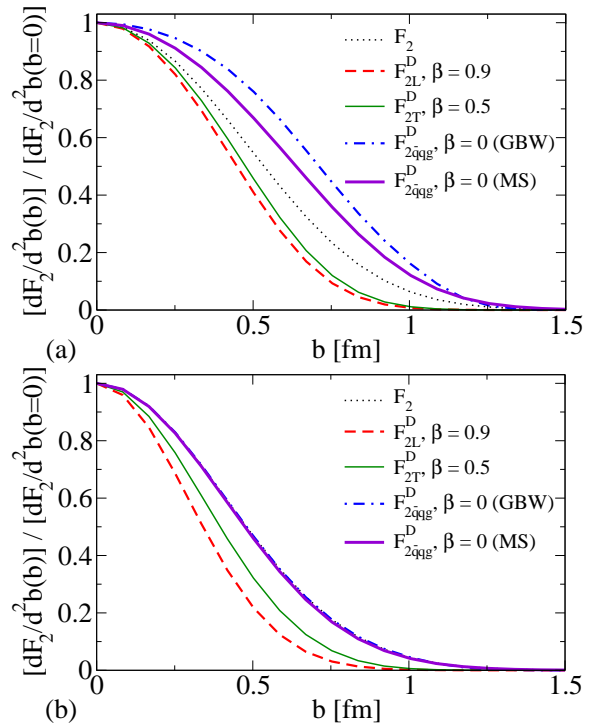

\includegraphics[width=0.45\textwidth]{dombetaprcv2.eps}
\includegraphics[width=0.45\textwidth]{dombetaq100prcv2.eps}
\caption{
$b$-dependence of the the inclusive structure function and
different contributions to the diffractive structure function at $Q^2 = 1 \gev^2$
(a) and  $Q^2 = 100 \gev^2$ (b) for
$x = 10^{-3}$ (inclusive) and $\xpom=10^{-3}$ (diffractive).
In plot (b) ($Q^2 = 100 \gev^2$) the $b$-dependence of 
the inclusive cross section and $q\bar{q}g$-components are indistinguishable.
}
\label{fig:btdep}
\end{figure}

We shall now discuss the $b$-dependence of the dipole cross-sections
(see also e.g. Refs.~\cite{Frankfurt:2001nt,Frankfurt:2005mc,Hatta:2007fg}).
Several works on the subject (for example,
Refs.~\cite{Golec-Biernat:1998js,Golec-Biernat:1999qd,%
Iancu:2003ge,Kugeratski:2005ck}) 
assume, explicitly or implicitly, a factorizable $\bt$ dependence
\begin{equation}\label{eq:factbt}
\dsigma(\bt,\rt,x) = 2\, \ampli(\bt,\rt,x) =  2 \,T_p(\bt) \amplibindep(\rt,x) \, .
\end{equation}
When considering diffractive scattering on protons, this is consistent with the 
exponential $t$
dependence observed in experiments, and in fact implies that $T_p(\bt)$ is Gaussian.

In the IPsat model, in contrast to the factorization of the $b$-dependence 
in \eq\nr{eq:factbt}, the dependence of 
the dipole cross-section on impact parameter is as in \eq\nr{eq:BEKW},
which equivalent to an impact parameter dependence of the saturation scale
$\qs^2 \propto T_p(\bt)$.
In the IPsat model
%, the impact parameter dependence of the saturation scale
% $\qs^2 \propto T_p(\bt)$ 
the impact parameter profile of the proton saturation scale is chosen to have the form 
\begin{equation}
T_p(\bt) = {1\over 2\pi B_{\rm G}}e^{-\frac{b^2}{2 B_{\rm G}}} \, ,
\label{bdep-IPsat}
\end{equation}
which is normalized to unity. 
In the large $Q^2$-limit the cross section is dominated by small dipoles and one can expand 
the exponential of \eq\nr{eq:BEKW}, and the dipole cross section becomes proportional 
to $T_p(\bt)$. This corresponds to $\langle b^2\rangle = 2 B_{\rm G}$, which can be 
interpreted as the average square gluonic radius in the proton.

 We shall now justify in the framework of the BK 
equation why we prefer~\cite{Kowalski:2003hm,Kowalski:2006hc,Kowalski:2007rw}
to incorporate the impact parameter dependence in the saturation scale $\qs$ instead of
 as a multiplicative factor. (As we shall see, this difference is more significant for 
diffractive as opposed to inclusive observables.) The fully impact parameter dependent BK
 equation for the dipole amplitude is
\begin{multline}\label{eq:BK}
\partial_y \ampli_{\xt,\yt}= \frac{\as N_c}{2\pi^2} \int \ud^2 \zt
\frac{(\xt-\yt)^2}{(\xt-\zt)^2(\yt-\zt)^2}
\\
\bigg[
\ampli_{\xt,\zt} + \ampli_{\zt,\yt} 
- \ampli_{\xt,\yt} -
\ampli_{\xt,\zt}\ampli_{\zt,\yt}
\bigg],
\end{multline} 
where $\bt = (\xt+\yt)/2$ and $\rt=\xt-\yt$.
This equation was studied numerically in Ref.~\cite{GolecBiernat:2003ym} with the 
conclusions, a) that the $\bt$ and $\rt$ dependence of the amplitude does not factorize,
 and b) one obtains unphysically large power law tails in $\bt$.
Physically this is due to the fact that the BK equation does not include 
confinement effects that would cut off these tails.
An approximation that removes the growth of the
power law tail with increasing $y=\log(1/x)$ is based on the argument that the dipole 
sizes $r'\sim 1/Q_s$ (which dominate the integration in \eqref{eq:BK}) are parametrically 
smaller than the typical scale $\rp$ for the variation of the amplitude with impact 
parameter $b.$ One can thus approximate (note that the same approximation was 
already implied when writing \eq\nr{eq:MS-amplitude})
\begin{multline}\label{eq:bindepBK}
\partial_y \ampli_{\bt,\rt}\approx \frac{\as N_c}{2\pi^2} \int \ud^2\rt'
\frac{\rt^2}{\rt'^2 (\rt-\rt')^2}
\\
\bigg[ 
\ampli\left(\bt,\rt'\right) +
 \ampli\left(\bt,\rt-\rt'\right)
- \ampli\left(\bt,\rt\right) -
\\
\ampli\left(\bt,\rt'\right)
\ampli\left(\bt,\rt-\rt'\right)
\bigg].
\end{multline} 
It is obvious that \emph{if} the $b$-independent equation has a scaling 
solution that can be expressed in terms of a saturation scale $\qs$, 
then replacing it with an impact parameter
dependent $\qs(\bt)$ gives a solution of \eq\nr{eq:bindepBK}.
This is effectively the approach used in 
Refs.~\cite{Lublinsky:2001yi,Levin:2001et,Gotsman:2002yy}.
The factorized ansatz \eq\nr{eq:factbt} is not (unless the profile is a $\theta$ function)
a solution of even this approximate equation. A factorized Gaussian profile for the 
proton dipole cross section, for example, \emph{does not approach the correct 
unitarity limit} for $b\neq0$. This is the main argument for including the impact parameter
dependence in the saturation scale, not as a factorizable prefactor of the
dipole cross section.

 Because $b$ is the Fourier 
conjugate variable of $\Delta$, where $t= -\Delta^2$ is the momentum 
transfer squared between the incoming and outgoing proton, a Gaussian dependence of the 
impact parameter profile corresponds to an exponentially decreasing dependence of the dipole
 cross-section on $t$. This exponential behavior is widely observed for diffractive final 
states~\footnote{
In contrast, the profile $T(\bt) = 2\theta(R-b),$ leads to the 
$t$-distribution~\cite{Frankfurt:2001av}
 $\sim 4 |J_1(\sqrt{-t}R)|^2/(-t R^2) \approx 1+tR^2/4+\mathcal{O}(t^2)$. }.
At smaller $Q^2$ (i.e. larger $r$) the impact parameter profile in the IPsat model
is not exactly Gaussian,
and diffractive peaks appear~\cite{Kowalski:2003hm} at large $-t$.
 This is, however, in a region
which is most likely unobservable because of the exponential suppression of the diffractive
cross section.
We emphasize (see also Ref.~\cite{Kowalski:2007rw}) that specifying the full
$\bt$-dependence of the dipole cross section 
simultaneously determines both the normalization (``$\sigma_0$'') of the cross section
 and the 
diffractive slope $B = \ud \ln \sigma^D/\ud t|_{t=0}$. 

In \fig\ref{fig:btdep}, we plot the inclusive and diffractive structure functions as a 
function of $b$, normalized to 
the same quantity at $b=0$. As has been noted previously, the inclusive distribution
 does not change significantly as $Q^2$ is varied. 
The $q{\bar q}g$ dipole contribution, at low $Q^2$, is significantly broader in
$b$ than the inclusive distribution. This should be interpreted as its
relative suppression at small $Q^2$ and small $b$, where the dipole cross section is 
close to the black disk limit and the $q{\bar q}g$ contribution vanishes, as 
is easily seen from \eq\nr{eq:MS-amplitude}.

\begin{figure*}
\begin{center}
\includegraphics[width=0.95\textwidth]{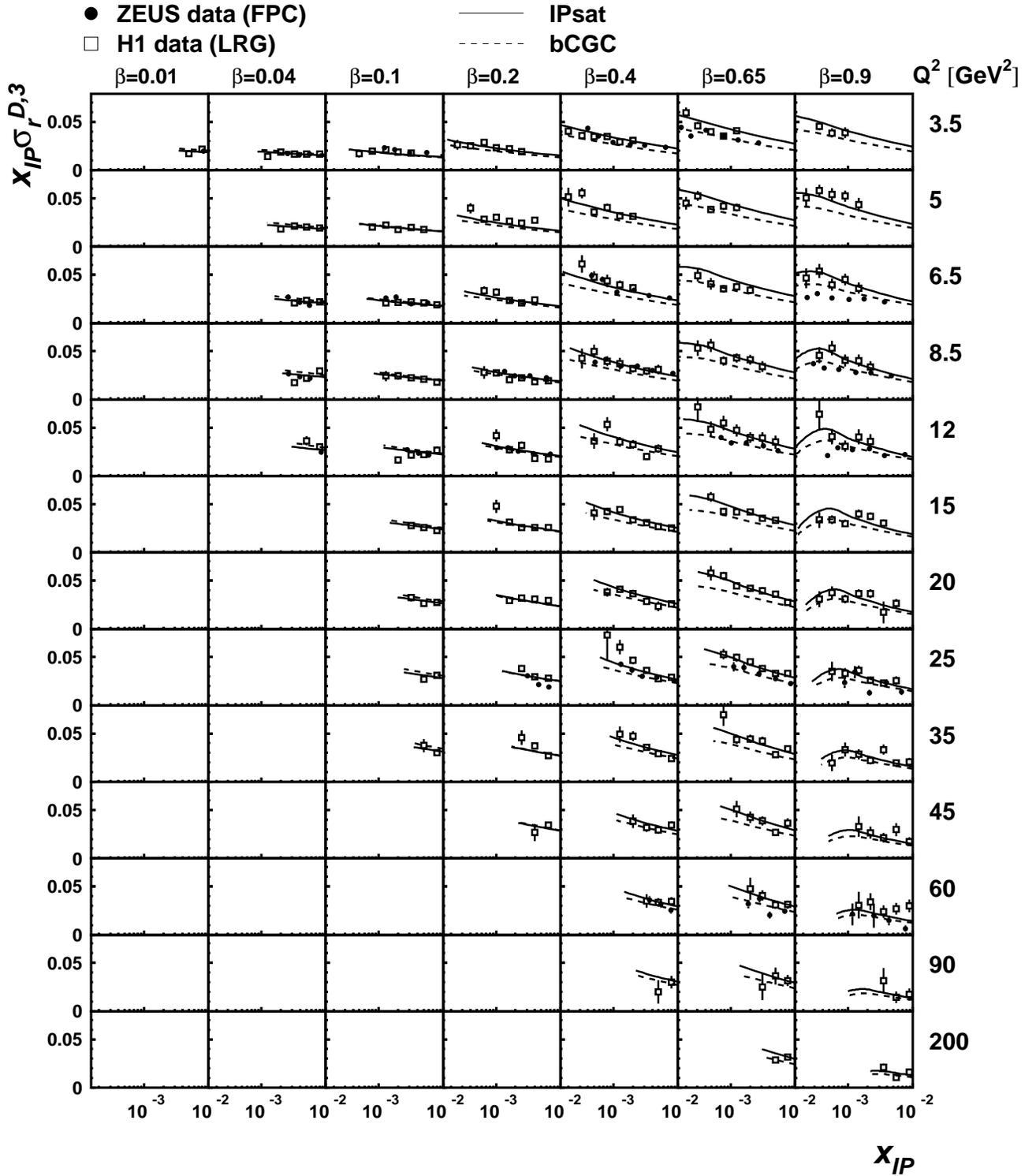}
\end{center}
\caption{
Comparison of the IPsat and bCGC fits to HERA data on diffractive structure functions.
}
\label{fig:hera}
\end{figure*}

Inclusive diffraction off nuclei is not very sensitive to the choice of impact parameter 
profiles in the proton. However, there are significant issues related to the impact 
parameter distribution of nuclei that are very relevant for diffraction, as we will 
see later. But before moving to scattering off nuclei we will first
discuss the comparison of the IPsat and bCGC
models to the HERA data on inclusive diffraction in ep scattering. 

\section{Comparison to HERA data}

An extensive comparison of the b-CGC and IPsat dipole models to the HERA data on 
{\it exclusive} vector meson production was performed in Ref.~\cite{Kowalski:2006hc}. 
We compare our calculation to the HERA results on diffractive structure functions
for $\xpom < 0.01$,
measured both using the rapidity gap method (ZEUS FPC \cite{Chekanov:2005vv} and
H1 LRG \cite{Aktas:2006hy}) and by measuring the recoil proton 
(ZEUS LPS \cite{Chekanov:2004hy} and H1 FPS \cite{Aktas:2006hx}).
Because the FPC and LRG data include events in which the proton has broken up, 
the cross-sections measured for the process $ep\!\rightarrow\!eXY$ are larger 
than the one measured for the process $ep\!\rightarrow\!eXp.$
We scale down this data by a constant factor to correct for the
proton dissociation contribution;  the ZEUS FPC data by a factor
of 1.45 and the H1 LRG data by 1.23. These factors are different due to the
different cuts on $M_Y$, the mass of the proton system.
Note that it is the FPS-LPS data that correspond to our definition of diffractive 
events and to our formulae, as the proton should escape the collision intact.
These correction factors were determined in the region 
$\beta \lesssim 0.7$ where data exists from both methods and then assumed to be 
independent of $\beta$; the largest contributions to the $\chi^2$ of our analysis
come from the large $\beta$ region where this assumption has not been tested
experimentally. All the $\chi^2$-values quoted below are calculated per degree of 
freedom using statistical and systematical errors added in quadrature. The total number 
of experimental points in the four data sets is 343, of which 76 are from the FPS-LPS 
data.

The experimental data are presented in terms of the reduced cross section 
$\sigma_r^{D,3}(\beta,\xpom,Q^2):$ 
\begin{equation}
\frac{d^3 \sigma^{ep\rightarrow eXp}}{d\xpom\ d\beta\ dQ^2}=\frac{4\pi\alpha_{em}^2}{\beta Q^4}
\left(1-y+\frac{y^2}{2}\right)\sigma_r^{D,3}(\beta,\xpom,Q^2)\ ,
\end{equation}
\begin{equation}
\sigma_r^{D,3}=F_T^{D,3}+\frac{2-2y}{2-2y+y^2}\ F_L^{D,3}\ .
\end{equation}
with $y\!=\!Q^2/(sx)$ where $\sqrt{s}\!=\!318\ \mbox{GeV}$ is the total energy in the 
$e\!-\!p$ collision.

With the IPsat cross sections, for the combined dataset from ZEUS and H1 data both with
 and without identified 
protons we get $\chi^2 = 1.3$ for $\as = 0.14$ in \eqs\nr{eq:gbw} and \nr{eq:ms}.
For the bCGC cross section the fit is equally good and works with a larger
value of the coupling: $\chi^2 = 1.3$ with $\as = 0.22$.
These are the values of $\as$ that we shall use for the respective models
to evaluate nuclear diffractive structure  functions in the next section.
For the IPsat model
the largest contribution to the $\chi^2$ comes from the rapidity gap method data at 
large $\beta$. The fit to only the LPS 
($\chi^2 = 0.5$ IPsat) and FPS ($\chi^2 = 0.8$) is much better. 
Considering just the LPS also accommodates a larger value of $\as=0.21$ with still 
$\chi^2<1$. The fit for bCGC is more
even among the datasets, but also there the H1 rapidity gap data has a larger
$\chi^2 = 1.9$ than the other data sets.
Our fit of these two models to a combination of the HERA datasets is presented in 
\fig\ref{fig:hera}.

The fit to HERA data is better with a smaller $\as$ than in Ref.~\cite{Marquet:2007nf}.
Given the $b$-dependence described previously this is to be expected. The factorized 
$b$-dependence (\eq\nr{eq:factbt}) used in earlier calculations of the diffractive
structure function  such as Refs.~\cite{Golec-Biernat:1999qd,Marquet:2007nf}
forces the $q\bar{q}g$-component to have the same impact parameter dependence as
the $q\bar{q}$-component. In a more realistic description, as suggested by \fig\ref{fig:btdep}, the $q\bar{q}g$ component
is sensitive to larger impact parameters and is thus larger; in order to fit the
same data this must be compensated by multiplying it with a smaller factor of $\as$.

\begin{figure}
\includegraphics[width=0.45\textwidth]{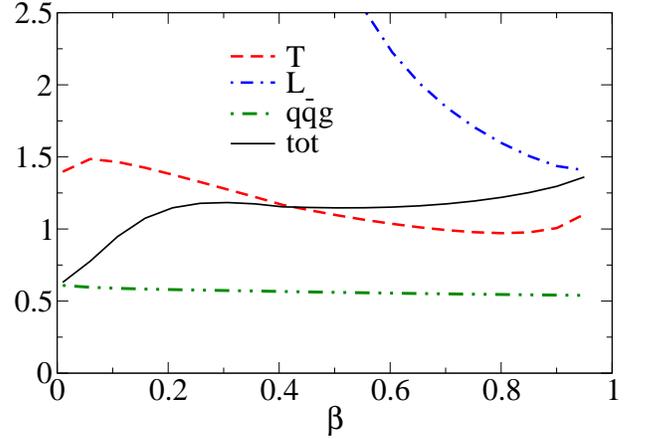}
\caption{
The ratio $\fda^x/(A \fdp^x)$ in the IPsat model for different components of the diffractive
structure function ($x=T,L,q\bar{q}g$) as a function of $\beta$ for gold
at $Q^2 = 5 \gev^2$ and $\xpom = 10^{-3}$ without nuclear breakup. 
}
\label{fig:diffshadcomps}
\end{figure}

\begin{figure}
\includegraphics[width=0.45\textwidth]{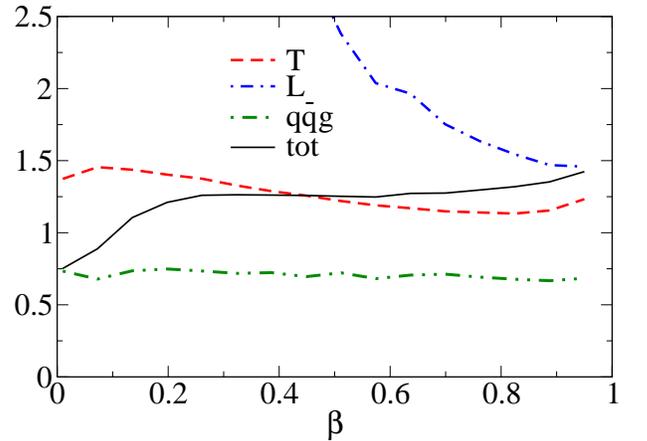}
\caption{
Same as Fig.~\ref{fig:diffshadcomps}, but including nuclear breakup.
}
\label{fig:diffshadcompsbup}
\end{figure}

\section{The nuclear diffractive structure function}

A straightforward generalization of the dipole formalism to nuclei
is to introduce the coordinates of the individual nucleons $\{\bt_i\}$. 
In the IPsat model the replacement 
$T_p(\bt) \to \sum_{i=1}^A T_p(\bt-\bt_i)$ gives the dipole cross section
\begin{equation}\label{eq:defsigmaa}
\dsigmaa
= 2 \left[1- 
e^{- r^2
F(\xbj,r) \sum_{i=1}^A T_p(\bt-\bt_i)
}
\right],
\end{equation}
where $F$ is defined in \eq\nr{eq:BEKW_F}. The positions of the nucleons
$\left\{\bt_i\right\}$
are distributed according to the Woods-Saxon distribution $T_A(\bt_i)$.
We denote the average of an observable $\mathcal{O}$
 over $\left\{\bt_i\right\}$ by 
 \begin{equation}
 \Aavg{\mathcal{O}} \equiv \int \prod_{i=1}^A \ud^2 \bt_i 
T_A(\bt_i) \mathcal{O}(\left\{\bt_i\right\}) \, .
\label{eq:nuc-average}
\end{equation}
Here we have introduced the Woods-Saxon thickness function
\begin{equation} \label{eq:defTA}
T_A(\bt) = \int \ud z \frac{C}{1+\exp \left[ 
\left( \sqrt{\bt^2 + z^2} - \ra \right)/d \right] },
\end{equation}
which is normalized to unity $\int \ud^2 \bt T_A(\bt) = 1$. The nuclear radius
$\ra$ and surface diffuseness $d$ are measured  from the electric charge 
distribution, their values can be found in Ref.~\cite{DeJager:1987qc}.
The average differential dipole cross section is well approximated
 by~\cite{Kowalski:2003hm} 
\begin{equation}
\Aavg{\dsigmaa} 
\approx 2\left[1-\left(1-\frac{T_A(\bt)}{2}\sigmap \right)^A\right]
\label{eq:nuc-dipole}
\end{equation}
where, for large $A$, the expression in parenthesis can be replaced by
form~\cite{Gotsman:1999vt}
\begin{equation}\label{eq:Glauber}
\exp\left(-\frac{A T_A(\bt)}{2}\sigmap \right).
\end{equation}
All parameters of the model come from either fits of the model to 
$ep$-data or from the Woods-Saxon distribution; no additional parameters are
 introduced for $eA$ collisions. 

The Glauber form \nr{eq:nuc-dipole} has a straightforward interpretation as the
dipole scattering independently off the different nucleons. To see this explicitly
denote $\dsigma(\rt,\bt) = 2(1-S(\rt,\bt)),$ where the $S$-matrix element $S(\rt,\bt)$
is the amplitude for the dipole to \emph{not} interact (elastically; the 
relation to the inclusive cross section is via the optical theorem) with the target. The
$S$-matrix element for scattering off a nucleus is then given by
$S_A(\rt,\bt) = \prod_{i=1}^A S_p(\rt,\bt-\bt_i)$ which, for the IPsat model,
turns out to be equivalent to $T_p(\bt) \to \sum_{i=1}^A T_p(\bt-\bt_i)$. 
Note that in the form \nr{eq:nuc-dipole} there is no leading twist shadowing, i.e. 
in the large $Q^2$ or small $r$ limit $\sigmaa \to A \sigmap$, 
because in this limit $\sigmap \sim r^2$ is small and one can expand the
exponential. 

The situation for the bCGC model is much more complicated, 
since the replacement $T_p(\bt) \to \sum_{i=1}^A T_p(\bt-\bt_i)$ into the definition 
of the bCGC saturation scale \nr{eq:qs-bCGC} does not lead to the Glauber 
form~\nr{eq:nuc-dipole}. One could see this as a consequence 
of the ``noncommutativity'' of nuclear effects and high energy evolution:
even if one assumes that for a particular $x$ and $\rt$ a dipole interacts
independently with the nucleons in a nucleus, this will not necessarily be
the case for other rapidities and dipole sizes because the evolution sums
up nonlinear interactions between the nucleons. Since it is not completely obvious 
how to introduce a nuclear dependence directly into the bCGC parametrization
for the dipole cross section we will in this work use \nr{eq:nuc-dipole}
for the bCGC model as well. A comparison of high energy evolution for protons and nuclei 
would be out of the scope of this work, see however 
Refs.~\cite{Mueller:2003bz,Albacete:2004gw}.

\begin{figure}
\includegraphics[width=0.45\textwidth]{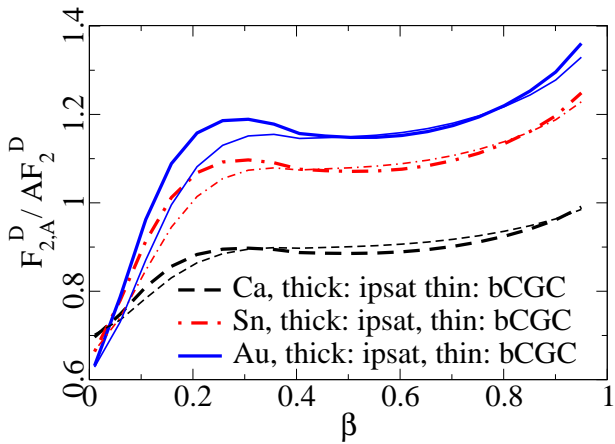}
\caption{
The ratio $\fda/(A \fdp)$ as a function of $\beta$ for Ca, Sn and Au nuclei for 
$Q^2 = 5$ GeV$^2$ and $\xpom = 10^{-3}$. Results are for the ``non breakup'' case 
in the IPsat model (thick lines)
and the bCGC model (thin lines).}
\label{fig:diffshadratiovsbeta}
\end{figure}

In Ref.~\cite{Kowalski:2007rw}, we showed that the nuclear 
dipole cross-sections 
obtained in this manner gave a good (parameter free) agreement with the $x$ 
and $Q^2$ dependence of the 
NMC inclusive structure function data~\cite{Amaudruz:1995tq,Arneodo:1996rv} 
at small $x$. However, at 
the level of the accuracy of the data, it 
was not possible to distinguish between the IPsat and b-CGC models for the
 inclusive cross-section. 
We will now consider nuclear diffractive ($q\bar{q}$ and $q\bar{q}g$) structure functions
 in the two dipole models. This is obtained by 
substituting the nuclear dipole cross-section (\eq\nr{eq:nuc-dipole}) in 
\eqs\nr{eq:phi-n}, \nr{eq:dip-adjoint} and \nr{eq:MS-amplitude}.

\begin{figure}
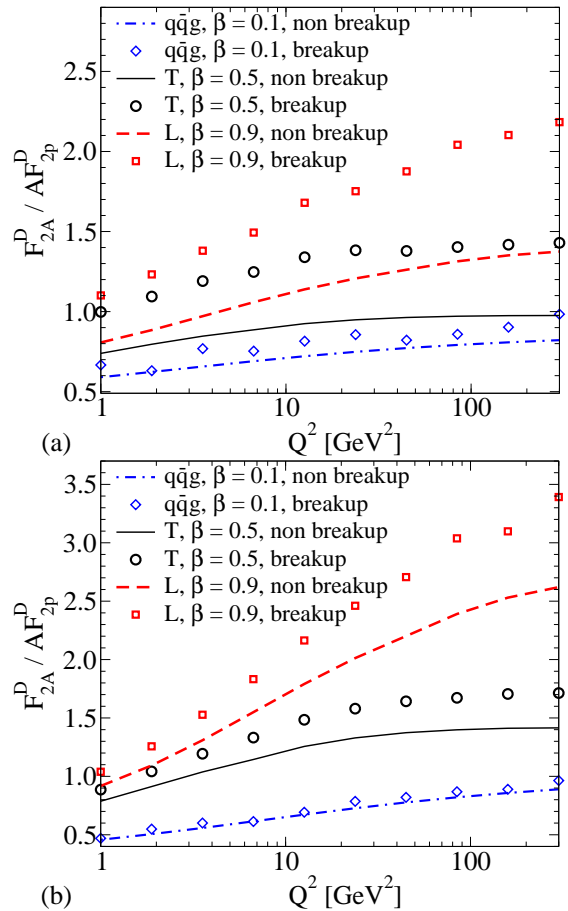

\includegraphics[width=0.45\textwidth]{diffshadvsqallcompscaipsatprcv2.eps}
\includegraphics[width=0.45\textwidth]{diffshadvsqallcompsauipsatprcv2.eps}
\caption{
The ratios $\fda^x/(A \fdp^x)$  at $\xpom = 10^{-3}$ for different components of the diffractive structure 
function plotted as a function of $Q^2$. The components are evaluated where they are
dominant: at $\beta=0.1$ for $q\bar{q}g$, $\beta = 0.5$ for $T$ and
$\beta = 0.9$ for $L$.
Results are in the IPsat model for  
both ``breakup'' and ``no breakup'' cases. (a) Ca nuclei, (b) Au nuclei.
}
\label{fig:diffshadratiovsQ2}
\end{figure}

It is very easy to break up a nucleus with a relatively small momentum transfer $|t| \gtrsim |t_{\rm min}^A|$. However, for 
$|t_{\rm min}^A| \lesssim |t| \lesssim |t_{\rm min}^p|$, where $t_{\rm min}^p$ is the minimum 
momentum transfer required to break up the proton, one can still have a nuclear diffractive
 event with a rapidity gap.  For $|t| \gtrsim |t_{\rm min}^A|$, the ``lumpiness'' of the 
nucleus shows up as a 
proton-like tail $\sim \exp\{C t\rp^2)\}$ of the $t$-distribution.
In our formalism, if one requires that the nucleus stays completely intact, the 
average $\Aavg{\cdot}$ in \eq\nr{eq:nuc-average} must be performed at the amplitude 
level; the results are therefore 
proportional to $\Aavgg{\dsigmaa}$. One finds that in this case  
$\ud\sigma^D/\ud t$ falls off very rapidly as $\sim \exp\{C t \ra^2\}$.
We refer to these 
(known as \emph{coherent} diffraction) as ``non breakup'' events.
 Measuring the 
intact recoil nucleus at such a small $t$ experimentally  
at a future electron ion collider~\cite{Deshpande:2005wd}
is challenging. Also including events where the nucleus breaks up into color neutral
constituents without 
filling  the rapidity gap between the $q\bar{q}$ dipole and the nuclear 
fragmentation region (\emph{incoherent} diffraction) 
corresponds to performing the average  $\Aavg{\cdot}$ over 
the cross section~\cite{Kovchegov:1999kx}; in this case, one performs the 
average $\Aavg{\left(\dsigmaa\right)^2}$ instead. The difference between 
the two averaging procedures can be significant with increasing values of
$t$ as shown explicitly in Ref.~\cite{Kowalski:2007rw}. 
The ``breakup'' figures in our plots are a sum of these
incoherent and coherent events.

In \figs\ref{fig:diffshadcomps} and \ref{fig:diffshadcompsbup}  we show the ratios 
of different components of the gold diffractive structure function to the proton one as a function of $\beta$. 
The $q\bar{q}$-components of the $\fda$ are enhanced compared to $A$
times the proton diffractive structure functions. This is to be expected, because of 
the fact that in a gold nucleus
the dipole cross section is, on average over the transverse area, closer to the 
unitarity limit than the proton - it is ``blacker''. The elastic scattering probability of a
$q\bar{q}$ dipole is maximal in the ``black'' limit and the approach 
to it is quicker in a large nucleus. 
The $q\bar{q}g$ component, on the other hand, is suppressed for nuclei compared to the proton. 
This is due to the fact that in a nucleus the scattering amplitude is closer to the unitarity
limit when the $q\bar{q}g$ component vanishes, as can be seen e.g. from \eq\nr{eq:MS-amplitude}.
 This leads to a nuclear suppression of the diffractive structure function in the 
small $\beta$ region,  where the $q\bar{q}g$ component dominates.
The net result of the different contributions is that $\fda$, for a large range in $\beta$,
is close to $A \fdp$.

\begin{figure}
\includegraphics[width=0.45\textwidth]{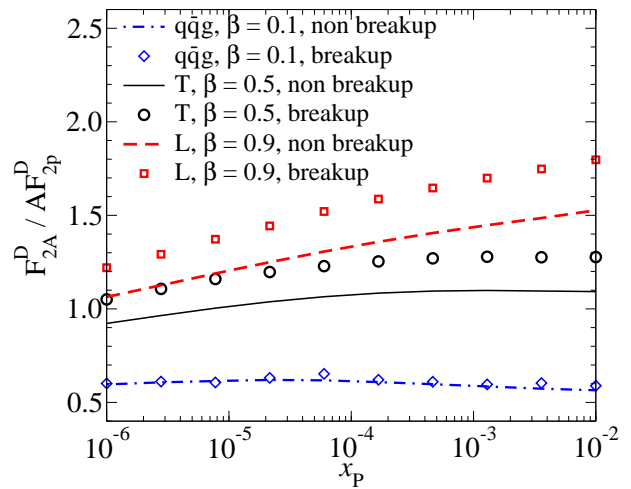}
\caption{
The ratios $\fda^x/(A \fdp^x)$ for different components of the diffractive structure 
function plotted as a function of $\xpom$. The components are evaluated at
$\beta=0.1$ for $q\bar{q}g$, $\beta = 0.5$ for $T$ and
$\beta = 0.9$ for $L$.
 Results are in the IPsat model for   Au nuclei 
both ``breakup'' and ``no breakup'' cases for $Q^2 = 5 \gev^2$.
}
\label{fig:ipsatvsxpom}
\end{figure}

\begin{figure}
\includegraphics[width=0.45\textwidth]{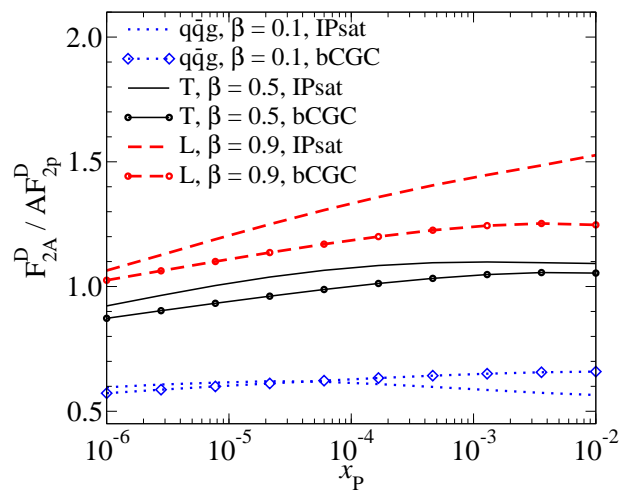}
\caption{
The ratios $\fda^x/(A \fdp^x)$ at $Q^2 = 5 \gev^2$ for the IPsat and bCGC models 
as a function of $\xpom$, for the ``no breakup'' case.
 The components are evaluated at
$\beta=0.1$ for $q\bar{q}g$, $\beta = 0.5$ for $T$ and
$\beta = 0.9$ for $L$.
}
\label{fig:ipsatbcgcvsxpom}
\end{figure}

\begin{figure}
 \includegraphics[width=0.45\textwidth]{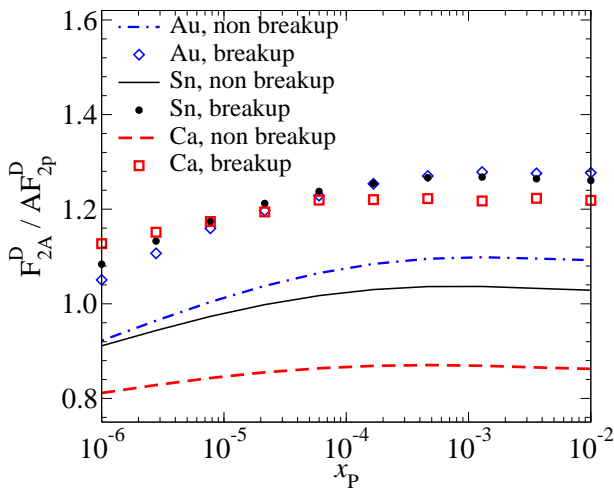}
\caption{
$F_{T,q\bar{q},A}^D/A F_{T,q\bar{q},p}^D $ at  $Q^2 = 5 \gev^2$ and $\beta=0.5$,
as a function of $\xpom$ for Au, Sn and Ca nuclei in the IPsat model.
 Both breakup and non-breakup cases are shown. 
}
\label{fig:ipsatTausnca}
\end{figure}

In \fig\ref{fig:diffshadratiovsbeta}, we plot the total ratio as a function of $\beta$ for 
different nuclei  in the ``non breakup'' case. As expected from our prior
 discussion, one sees a strong enhancement with $A$ for larger $\beta$ and likewise, a 
stronger suppression with $A$ at very small values of $\beta$. A comparison of the ``breakup'' 
versus ``non breakup'' cross-sections can be seen in \fig\ref{fig:diffshadratiovsQ2} for the 
ratio of 
diffractive cross-sections as a function of $Q^2$.
The results in \fig\ref{fig:diffshadratiovsQ2} for the ratio of diffractive structure 
functions indicate that the diffractive cross-section in nuclei 
decrease more slowly for large $Q^2$ than in the proton.

In \figs\ref{fig:ipsatvsxpom}, \ref{fig:ipsatbcgcvsxpom} and \ref{fig:ipsatTausnca}
we show the $\xpom$ dependence of the nuclear modifications. For a fixed $Q^2 = 5\gev^2$
the nuclear enhancement of the $q\bar{q}$ components becomes smaller at smaller $\xpom$.
This can easily be understood as an analogous effect to the $Q^2$-dependence plotted in 
\fig\ref{fig:diffshadratiovsQ2}: increasing $\xpom$ for a fixed $Q^2$ increases 
$Q^2/\qs^2(\xpom)$ and has the same effect as increasing $Q^2$ at a fixed $\xpom$.
In \fig\ref{fig:ipsatbcgcvsxpom} we compare the $\xpom$-dependence in the IPsat and 
bCGC models. As was already observed in \cite{Kowalski:2007rw}, the experimental
signature for the different evolution dynamics in the models in eA-scattering is 
mainly in the different $\xpom$ dependence in the nuclear modification factor. 
The result presented in \fig\ref{fig:ipsatbcgcvsxpom} confirm this, although the effect
is perhaps smaller than in the shadowing of the inclusive cross section. 
In \fig\ref{fig:ipsatTausnca} we compare, in the IPsat model, the nuclear modifications to 
$F_{T,q\bar{q}}^D$ at $\beta = 0.5$ for Ca, Sn and Au nuclei. One can see that the 
``non breakup'' curves are much more sensitive to the nuclear species than the 
``breakup'' ones.

\begin{figure}
 \includegraphics[width=0.45\textwidth]{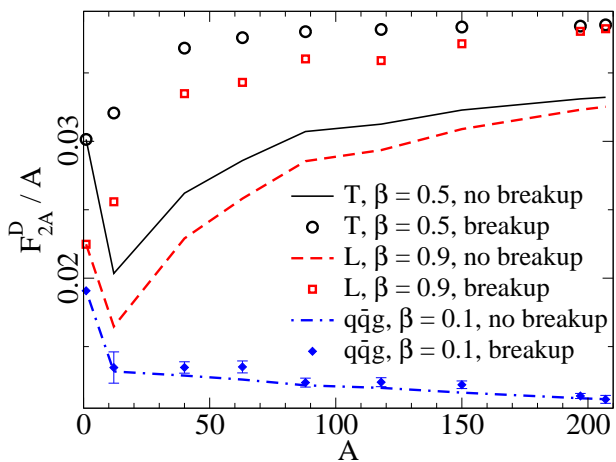}
\caption{
$F_{T,q\bar{q},A}^D/A$ at $\beta=0.5$,
$F_{L,q\bar{q},A}^D/A$ at $\beta=0.9$ 
and $F_{q\bar{q}g,A}^D/A$ at $\beta=0.1$
vs. A at $Q^2 = 5 GeV^2$, $\xpom = 10^{-3}$.
 Both breakup and non-breakup cases are shown. 
}
\label{fig:diffshadratiovsA}
\end{figure}

In 
\fig\ref{fig:diffshadratiovsA}, the dependence of the longitudinal 
and transverse components of the diffractive structure function on nuclear size is 
shown for the ``breakup'' and ``non breakup'' cases. In the ``breakup'' case, one 
sees a very weak $A$ dependence
In the coherent ``non breakup'' case, one first notes that the 
diffractive structure function first decreases up to atomic numbers $A\sim 10$, 
before beginning to rise.  As noted in 
Ref.~\cite{Kowalski:2007rw}, this is due to the typical scattering amplitude
for small nuclei actually being smaller than for a proton because of the diluteness 
of the nucleus. This leads to a suppression of coherent diffraction. 
The ``breakup'' case, on the other hand, can only be enhanced in nuclei.
For gold nuclei, the cross sections in the 
``non breakup'' case are about $15$\% lower than in the ``breakup'' case.

Because of the different nuclear modifications in inclusive and diffractive scattering,
the fraction of diffractive events in an experiment
depends on the detailed kinematics and experimental coverage.
Thus it is not straightforward to give a very precise general estimate for 
$\sd/\stot$  that would be observed in a generic high energy eA collider.
A general order of magnitude argument would be as follows. For moderate values of
$Q^2$ and large nuclei we expect a nuclear shadowing of the inclusive structure
function by a factor $\sim 0.8$~\cite{Kowalski:2007rw}. A typical nuclear 
enhancement of diffraction (at moderate values of $\beta \gtrsim 0.2$)
is a factor of $\sim 1.2$ (see e.g. \fig\ref{fig:diffshadratiovsbeta}). Combining
these we expect $\sd/\stot$ to be increased by a factor of $1.2/0.8 = 1.5$ compared to the 
proton. Thus from a typical ep fraction of 15\% we expect $\sd/\stot$ to go up 
to 20\% -- 25\% at an eA collider.

\section{Comparisons and Conclusions}

Finally we shall briefly compare our results to some other recent literature on 
diffraction in eA scattering. Several aspects of the treatment of nuclear diffraction here 
were discussed previously in the works of Nikolaev, Zakharov and 
Zoller~\cite{Nikolaev:1995xu}  and of 
Frankfurt and Strikman~\cite{Frankfurt:1991nx} --- these works 
may be consulted for earlier works in the literature as well. 
With regard to these early works, a key difference is the use of the explicit 
form of the $q{\bar q}g$ contributions to the diffractive structure functions
 given in \eq\nr{eq:gbw} and \nr{eq:ms}. Further, we have focused on the predictions
 of Color Glass Condensate based models, that were previously constrained from fits
 to HERA data. In this paper, we have made explicit fits of these models to the HERA 
diffractive structure function data as well for the first time.

Turning to relatively more recent works in the literature, Gotsman, Levin, Lublinsky,
 Maor and Tuchin~\cite{Gotsman:1999vt}
use a parametrization of the impact parameter dependence that is similar to ours 
to compute the fraction of diffractive events in the total cross section. They see a 
modest enhancement compared to ep scattering in agreement with our results. As
 previously mentioned, they use a different method to calculate the $q\bar{q}g$ 
contribution
but their results (see for example, Fig.~7 in Ref.~\cite{Gotsman:1999vt}) seem to point
towards its relative suppression in nuclei, in qualitative agreement with ours. On 
the other hand the energy dependence seems much stronger than our results indicate. 
Since the results of  Ref.~\cite{Gotsman:1999vt} are not presented in the form 
of diffractive structure functions (in particular the $\beta$-dependence is not 
calculated) a more detailed comparison is difficult. In Ref.~\cite{Levin:2002fj} 
Levin and Lublinsky start from a Glauber-like parametrization
very similar to our \eq\nr{eq:Glauber}, but end up with a saturation scale depending on 
the nuclear size as $\qs^2 \sim A^{0.6}$
for moderately small values of $x\sim 10^{-3}$, which naturally leads to a much stronger 
$A$-dependence of $\sigma_D/\sigma_{\mathrm{tot}}$ than our results.

In the work of Frankfurt, Guzey and 
Strikman~\cite{Frankfurt:2001av,Frankfurt:2002kd,Frankfurt:2003gx}, 
the nuclear diffractive structure functions are 
modelled starting from leading twist diffractive parton distributions in 
the proton. Although the terminology and theoretical framework are somewhat different, 
some comparisons can still be made. They find that $\fda/A$ is suppressed compared
to $\fdp$, as a function of $Q^2$,  both at small $\beta = 0.1$ (in agreement with 
our result)
and in the transverse $q\bar{q}$ dominated region $\beta = 0.5$
(which disagrees with our findings); this can be seen explicitly by comparing
 \fig5 in \cite{Frankfurt:2003gx} and our \fig\ref{fig:diffshadratiovsQ2}.

Kugeratski, Goncalves and  Navarra \cite{Kugeratski:2005ck,Goncalves:2006yt,Cazaroto:2008iy}
compute nuclear diffractive structure functions using 
the GBW framework~\cite{Golec-Biernat:1999qd}, albeit with the 
IIM~\cite{Iancu:2003ge} dipole cross section. They extend the
calculation to nuclei using a simple $A^{1/3}$-scaling of the saturation scale, the
transverse area $\sigma_0$ and the diffractive slope $B_D$; the latter two are treated 
as independent parameters. They observe (see \figs4 and 6 in 
\cite{Kugeratski:2005ck}) a similar pattern in the $\beta$-dependence
as we do, namely that the nuclear $\fd$ is relatively suppressed at small $\beta$.
However, from their results one also deduces 
(note that the diffractive structure functions in \cite{Kugeratski:2005ck} are 
normalized by $A^{4/3}$ which is not explicit in the text~\cite{gonapriv})
that $\fda/A \fdp$ is of the order of $1/4$,
which is much smaller than our result 
$\sim 1$ (see \fig\ref{fig:diffshadratiovsbeta}).  We conjecture that these
 differences arise because, unlike in our discussion, $\sigma_0$ and 
$B_D$ are treated as independent parameters in Ref.~\cite{Kugeratski:2005ck}, with
 values that are not necessarily consistent with each other.

Measuring diffractive events in a high energy eA collider would yield significant
insight into the physics of high parton densities. The experimental measurements
of inclusive diffraction are encoded in diffractive structure functions that
depend on the kinematical variables of the process. We have argued
that it is possible to make a large number of detailed theory predictions
for nuclear diffractive structure functions. Comparing these to experimental measurements
would provide a powerful handle to understanding the behavior of QCD at high energy.
In this paper we have presented some of these theory predictions, within two particular
models for the high energy wavefunction. One (IPsat) is constructed to combine the
correct DGLAP limit at large $Q^2$ with unitarization at small $Q^2$. The other
(bCGC) is a parametrization of the leading order BK dynamics.
We have also emphasized the importance of including in the calculations a 
realistic description of the impact parameter dependence in
the nucleus, without which an actual comparison to experimental data is not very useful.

\section*{Acknowledgments}

We thank M. Strikman, M.~S. Kugeratski, V.~P. Goncalves and F.~S. Navarra 
for discussions.
RV's research is supported by DOE Contract No. DE-AC02-98CH10886 and
CM's research is supported by the European Commission under the FP6
 program, contract No. MOIF-CT-2006-039860.

\bibliographystyle{JHEP-2mod}
\bibliography{spires}

\end{document}